\documentclass[prb,aps,twocolumn,showpacs,preprintnumbers,amsmath,amssymb,floatfix]{revtex4-2}
\usepackage{amsmath}
\usepackage{graphicx}
\usepackage{dcolumn}
\usepackage{bm}
\usepackage{float}
\usepackage{xcolor}
\usepackage[colorlinks=true,linkcolor=blue]{hyperref}

\begin{document}

\title{Edge-Aware Graph Attention Model for Structural Optimization of High Entropy Carbides}

\author{Neethu Mohan Mangalassery}
\author{Abhishek Kumar Singh}
\email{abhishek@iisc.ac.in}
\affiliation{Materials Research Centre, Indian Institute of Science, Bangalore 560012, India}

\begin{abstract}
Predicting relaxed atomic structures of chemically complex materials remains a major computational challenge, particularly for high-entropy systems where traditional first-principles methods become prohibitively expensive.
We introduce the edge-aware graph attention model, a physics-informed graph neural network tailored for predicting relaxed atomic structures of high-entropy systems. the edge-aware graph attention model employs chemically and geometrically informed descriptors that capture both atomic properties and local structural environments. To effectively capture atomic interactions, our model integrates a multi-head self-attention mechanism that adaptively weighs neighbouring atoms using both node and edge features. This edge-aware attention framework learn complex chemical and structural relationships independent of global orientation or position. We trained and evaluated the edge-aware GAT model on a dataset of carbide systems, spanning binary to high-entropy carbide compositions, and demonstrated its accuracy, convergence efficiency, and transferability. The architecture is lightweight, with a very low computational footprint, making it highly suitable for large-scale materials screening. By providing invariance to rigid-body transformations and leveraging domain-informed attention mechanisms, our model delivers a fast, scalable, and cost-effective alternative to DFT, enabling accelerated discovery and screening of entropy-stabilised materials.
\end{abstract}

\maketitle

\section{Introduction}

High-entropy materials (HEMs)~\cite{schweidler2024high,ren2025review}, including high-entropy alloys (HEAs)~\cite{praveen2018high,wang2021high,miracle2014exploration} and carbides (HECs)~\cite{fan2024high,sarker2018high}, have emerged as next-generation materials due to their remarkable mechanical strength, thermal stability, and resistance to radiation damage and corrosion. These exceptional properties make HEMs highly attractive for aerospace, nuclear reactors, cutting tools, protective coatings, and other extreme environment applications.

HEAs consist of multiple principal elements in near-equiatomic ratios, offering a unique combination of strength, ductility, and phase stability, making them suitable for high-performance structural components such as turbine blades and engine  \cite{miracle2014exploration}. HECs, on the other hand, combine the entropy-driven stability of HEAs with the intrinsic hardness of carbides, producing ultra-hard, wear-resistant materials ideal for nuclear fuel cladding, thermal barrier coatings, and machining applications \cite{feltrin2022review}.

Designing such advanced materials requires accurate prediction and optimization of atomic structures, as atomic configurations govern fundamental properties like hardness, thermal conductivity, chemical stability, and electronic behavior. Reliable structural modeling is therefore critical across fields including catalysis, energy storage, additive manufacturing, and high-temperature structural components.

Despite their promise, rational design of HEMs remains challenging due to the vast compositional space and complex atomic-scale disorder. Traditional first-principles methods, such as density functional theory (DFT)~\cite{orio2009density,geerlings2003conceptual}, provide high accuracy but are computationally prohibitive for large, chemically complex systems.

To address this challenge, machine learning interatomic potentials (MLIPs)~\cite{mishin2021machine,kulichenko2024data,novikov2020mlip} have emerged as faster alternatives. However, conventional MLIPs often rely on handcrafted descriptors and are limited to narrow chemical domains, restricting their ability to generalize across different materials. Moreover, they primarily capture short-range interactions and may overlook long-range or many-body effects, which are critical in HEMs with highly varied atomic environments.

Recent deep learning approaches have enabled models that learn atomic interactions directly from structural and chemical data. For instance, DeepRelax~\cite{yang2024scalable} employs a generative model for crystal structure relaxation without iterative energy minimization. While effective for many systems, its dependence on local descriptors, large DFT datasets, and high computational cost limits performance for chemically disordered systems such as HEAs and HECs. Other methods, such as graph neural networks with data augmentation~\cite{gibson2022data} or GAN-based translation models~\cite{kim2023structure}, improve predictions but struggle to generalize to compositionally complex high-entropy systems or require extensive training data. These approaches, though promising, often act as black boxes and demand significant computational resources.

In this work, we introduce an edge-aware graph attention model (edge-aware GAT), a deep learning framework specifically designed to predict relaxed atomic positions in chemically complex materials. Our model incorporates edge-aware attention mechanisms and chemically informed descriptors inspired by the CLEAR (Chemistry and Local Environment Adaptive Representation) framework~\cite{swetlana2024chemistry}. CLEAR descriptors can encode both chemical identity and geometric context through Voronoi tessellation. We used modified CLEAR descriptors with additional edge features, including scalar interatomic distances, distance vectors, unit direction vectors, and mean bond angles, while node features capture elemental properties and local atomic concentration.
The edge-aware GAT learns complex interatomic dependencies through iterative message passing, capturing both short-range and long-range interactions essential for structural relaxation. The resulting unified graph representation combines positional, compositional, and geometric features, enhancing generalizability across both ordered and disordered material systems.

We trained the model on a curated dataset of 925 carbide structures, spanning binary carbide compounds to quaternary carbide systems containing 12 transition metals and carbon. The model achieves a mean absolute error of 0.09~\AA~in predicting relaxed atomic positions, demonstrating its effectiveness in modelling structural relaxation across chemically diverse systems. By reducing reliance on costly DFT calculations and handcrafted descriptors, the proposed framework provides an efficient tool for accelerated materials discovery and detailed structure–property analysis across chemically diverse systems.

\section{Methodology}

\subsection{Theoretical calculations}
The random solid solutions of FCC HECs are build using special quasi-random structures (SQS) of ATAT package\cite{zunger1990special}. The HECs are built using a 64-atom supercell with the size of (2 X 2 X 2) as the parent structure. The relaxed structures of FCC HECs are calculated using first-principle density functional theory (DFT) with projector-augmented wave (PAW) potentials as implemented in the Vienna Ab Initio Simulation Package (VASP)\cite{kresse1996efficient,kresse1996efficiency,blochl1994projector}. \par All-electron PAW potentials describe electron-ion interactions. The electronic exchange and correlation potential terms are represented by the Perdew-Burke-Ernzerhof (PBE) generalised gradient approximation (GGA)\cite{perdew1996generalized}. The energy cut-off of 400eV was used to expand the plane-wave basis sets. All the HEC structures are fully relaxed using the conjugate gradient method until the Hellmann-Feynman forces on every
atom is less than 0.05 eV/Å. A well-converged gamma centered k-grid
of 3 × 3 × 3 is used to sample the Brillouin zone (BZ), and the Methfessel-Paxton scheme with a smearing width of 0.1 eV is employed\cite{methfessel1989high}. 
\begin{figure*}
    \centering
    \vspace{2mm}
    \includegraphics[width=1\linewidth]{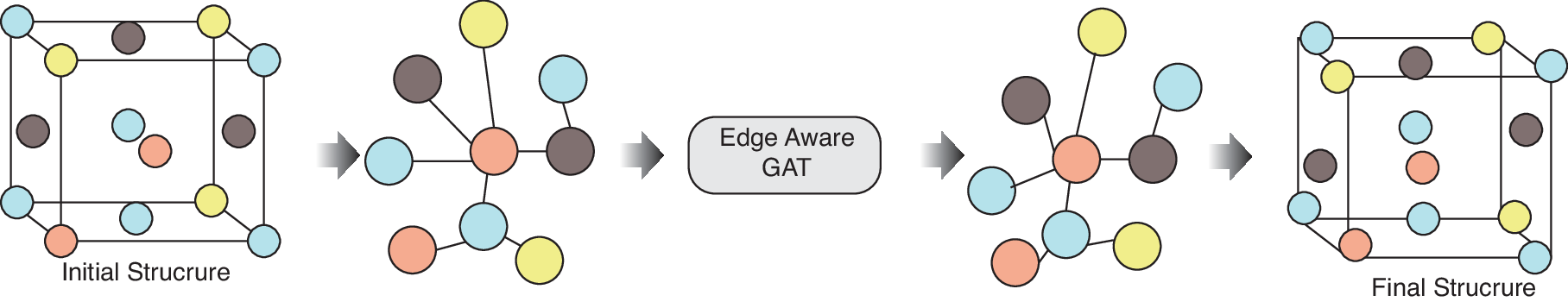}
    \caption{Overall workflow of the proposed Edge-Aware GAT model from the initial structure to the final refined structure.}
    \label{fig1}
    \vspace{-2mm}
\end{figure*}
    \subsection{GNN and GAT}
    Graph Neural Networks (GNNs)~\cite{corso2024graph,xie2018crystal,reiser2022graph,korolev2019graph} have emerged as powerful tools in materials science for modeling atomic systems by capturing both local and non-local atomic interactions. Among the various GNN architectures, attention-based models such as Graph Attention Networks (GATs)~\cite{vrahatis2024graph,xie2020mgat,schmidt2021crystal,banik2023cegann} are particularly well-suited for this task, as they learn to dynamically weight the influence of neighboring atoms. Building upon this, we develop a dynamic edge-aware graph attention network for high entropy systems~\cite{mo2022multi,banik2023cegann,zhang2022eesanet}.This is designed to predict relaxed atomic positions from unrelaxed crystal structures using a graph-based deep learning framework.
    
    \section{Result and Discussion}
\subsection{Dataset Preparation and Feature Extraction}

To systematically investigate the structural behavior of FCC high-entropy carbides, a pool of 12 transition metals together with carbon was selected for structure generation using the SQS method. A total of 925 unique configurations were generated, each containing 64 atoms. These structures span a broad compositional space, including 12 MC carbides, 198 M\textsubscript{1}M\textsubscript{2}C carbides, 210 M\textsubscript{1}M\textsubscript{2}M\textsubscript{3}C carbides, and 495 quaternary M\textsubscript{1}M\textsubscript{2}M\textsubscript{3}M\textsubscript{4}C carbides, thereby covering all possible combinations of the selected elements. From this dataset, 740 structures were used for training and 185 structures were reserved for testing, ensuring a diverse range of chemical and structural environments accessible to the model.

\par Node feature preparation focused on constructing a chemically meaningful representation for each atom in the system. Every atom was assigned a feature vector composed of 27 elemental descriptors, including physicochemical, structural, and electronic properties that are relevant for capturing trends across transition-metal carbides. In addition, atomic concentration information obtained from the SQS configurations was incorporated so that the node features encode both local elemental identity and global composition. Because these features depend solely on intrinsic elemental properties and composition, not on absolute spatial coordinates—they remain invariant under global translations of the structure. These node embeddings form the initial atomic representation used by the model before any geometric information is incorporated.

\par Edge feature preparation involved characterizing the geometric and chemical relationships between atomic pairs $(i,j)$ identified within a cutoff radius or through Voronoi connectivity. The edge descriptor combines scalar, angular, and vectorial components and is formally expressed as:
\begin{equation}
    \mathrm{EdgeFeatures(e}_{ij}) = \left[ d_{ij},\; (c_i a_i - c_j a_j),\; \bar{\theta}_{ij},\; \mathbf{d}_{ij},\; \hat{\mathbf{r}}_{ij} \right],
\end{equation}
where $d_{ij}$ is the interatomic distance, $(c_i a_i - c_j a_j)$ encodes concentration weighted differences in elemental descriptors, $\bar{\theta}_{ij}$ is the average angle between the $i$--$j$ bond and bonds from atom $i$ to its neighboring atoms, $\mathbf{d}_{ij}$ is the displacement vector from $i$ to $j$, and $\hat{\mathbf{r}}_{ij}$ is the corresponding unit vector. 
The node and edge features were inspired by the CLEAR descriptor \cite{swetlana2024chemistry}, which encodes atomic environments using concentration-weighted elemental properties and geometric information. In the original CLEAR formulation, the descriptors are often expressed in terms of statistical summaries, such as mean and standard deviation of pairwise atomic differences over the neighborhood.

For the present work, we adopt a modified version of the CLEAR descriptor tailored for predicting relaxed atomic positions. Specifically, we treat each atomic pair $(i,j)$ separately rather than aggregating statistics, and extract the following components for edge features: the Euclidean bond length $d_{ij}$ and the concentration-weighted elemental difference $c_i a_i - c_j a_j$. By retaining individual pairwise information, our modified descriptor captures fine-grained local atomic interactions.

\begin{figure*}[!t]
    \centering   
    \vspace*{-5mm}
    \includegraphics[width=1\textwidth]{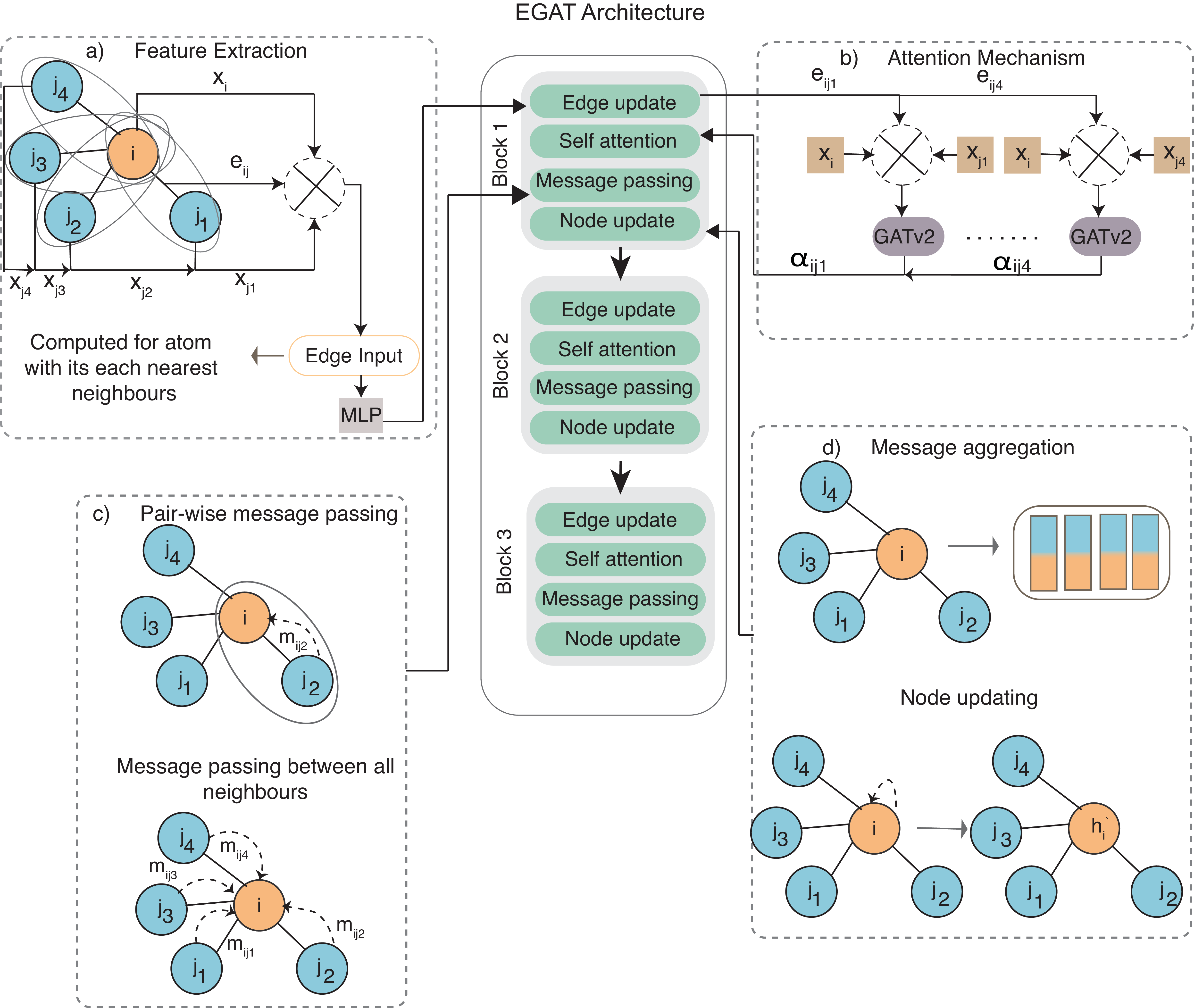}
    \caption{Architecture Edge-aware GAT model. Each carbide system is represented as a crystal graph, where atoms are nodes and bonds are edges. Node features include elemental features and composition, edge attributes contain structural and chemical features. These descriptors are updated via an MLP will repeat in three blocks, and an attention mechanism selects the most influential neighbors for message passing. Final attention coefficients are used to update node features.}
    \label{fig2}
    \end{figure*}

Scalar quantities such as $d_{ij}$ and $\bar{\theta}_{ij}$ are inherently invariant to global translations and rotations because they depend only on relative positions and angles between atoms. For instance, if the entire crystal is shifted in space or rotated, the distance between two atoms and the angles formed with neighboring atoms remain unchanged. This invariance ensures that the predictions of model are not affected by the arbitrary positioning or orientation of the structure.

Vectorial quantities, including $\mathbf{d}_{ij}$ and $\hat{\mathbf{r}}_{ij}$, encode directional information that is essential for representing the 3D geometry of the atomic environment. These vectors transform equivariant under rotation, meaning that if the entire structure is rotated, the vectors rotate in the same manner. This property allows the model to preserve directional relationships between atoms, which is critical for accurately reconstructing atomic displacements and for capturing orientation-dependent interactions, while still respecting the symmetry of the system.

By combining these invariant and equivariant components, the edge features provide a complete, symmetry-aware description of atomic interactions: scalar features guarantee stability under global transformations, and vector features retain orientation information. This dual property allows the network to learn meaningful structural patterns without being misled by arbitrary rotations or translations of the system.

\par During feature extraction, the atomic positions of both unrelaxed and relaxed structures were stored as \(N \times 3\) coordinate matrices, capturing the complete spatial arrangement of atoms. The structural data were then converted into graph form by applying Voronoi tessellation to identify the nearest neighbors of each atom. Each atom was treated as a graph node, and edges were created between Voronoi-adjacent atoms, ensuring that the constructed graph reflects the physically meaningful coordination environment. Once the graph connectivity was established, the previously defined node and edge descriptors were computed for all structures. This unified workflow from dataset construction to feature computation provides a consistent and symmetry-aware representation of chemical and geometric environments, enabling the model to learn the structural patterns required for accurate prediction of relaxed atomic positions.
\par The overall workflow of the present study, from crystal structure generation to relaxed position prediction, is schematically illustrated in Fig.~\ref{fig1}. Starting from the initial unrelaxed atomic structures, the systems are first converted into graph representations through Voronoi-based neighbor construction. Subsequently, the extracted node and edge features are supplied to the proposed edge-aware graph attention network (EGAT), which learns the complex chemical and geometric interactions and predicts the relaxed atomic positions.
\subsection{Edge-aware GAT Model Architecture}

The overall workflow of the proposed edge-aware graph attention (EGAT) model is illustrated in Fig.~\ref{fig2}. For each input carbide structure, the first step involves constructing a crystal graph representation, where atoms are treated as nodes and the geometric connectivity between atoms is encoded through edges. The local atomic neighborhood for each node is identified using the Voronoi tessellation method, which enables a physically meaningful determination of nearest neighbors without relying on fixed cutoff distances.

Each node is initialized with a 28-dimensional feature vector formed by concatenating a 27-dimensional elemental property descriptor with the corresponding atomic concentration obtained from the SQS-generated configuration. These features provide a chemically informed description of each atom in the crystal. For every edge connecting a pair of neighboring atoms, a 35-dimensional feature vector is constructed to encode both geometric and chemical information of the atomic pair. The complete list and detailed description of all node and edge features are provided in the Supplementary Information.

Once the crystal graph structure and all associated node and edge features are fully prepared, the resulting graph is passed as input to the edge-aware graph attention network. Prior to entering the edge aware GAT layers, all features are standardized to ensure numerical stability and balanced learning. Within the layers, both node and edge features are jointly processed through attention-based message passing, Figure~\ref{fig2} presents a schematic overview of the proposed edge-aware GAT (EGAT) architecture and its main computational stages. Panel (a) illustrates the initial feature extraction and edge initialization process, where each atom is connected to its nearest neighbors identified through the Voronoi construction. The pre-computed node features and raw edge features obtained from geometric and chemical descriptors are combined and passed through an initial edge update network (MLP) to generate the input edge embeddings for the EGAT model. 
Panel (b) depicts the attention mechanism based on the GATv2 formulation, where attention coefficients are computed using transformed node features together with the updated edge features, enabling adaptive and edge-aware neighbor weighting. Panel (c) shows the pair-wise message passing process, in which information is exchanged between a central atom and all its neighboring atoms through the learned attention coefficients and refined edge representations. Panel (d) illustrates the message aggregation and node updating step, where messages from all neighbors are aggregated and used to update the central node representation. 
These four stages are stacked into multiple EGAT blocks, enabling progressive refinement of both node and edge features across the network. Each component is described in detail in the following subsections.
Three edge-aware GAT blocks are stacked sequentially, progressively refining node and edge representations. After the final block, the resulting node embeddings are mapped through a linear output layer to predict the relaxed atomic coordinates. The model is trained by minimising the mean absolute error (MAE) between predicted and DFT-relaxed coordinates, using the Adam optimizer with weight decay to ensure stability and convergence.

\subsubsection{Edge Update Block}

As illustrated in Fig.~2(a), the edge update block serves as the first learnable operation within each EGAT layer and is applied immediately after graph construction and feature extraction. Its role is to refine the raw edge descriptors using the current node and edge representations before attention-based message passing is performed.

For each edge connecting a source atom $i$ and a destination atom $j$, the input to the edge update block is constructed by concatenating the corresponding source node features $\mathbf{x}_i$, destination node features $\mathbf{x}_j$, and the existing edge features $\mathbf{e}_{ij}$:
\[
\mathbf{z}_{ij} = [\mathbf{x}_i \, \| \, \mathbf{x}_j \, \| \, \mathbf{e}_{ij}].
\]
This combined representation is passed through a multi-layer perceptron (MLP) consisting of successive linear transformations followed by LeakyReLU activations and batch normalization. The MLP outputs a correction term $\Delta \mathbf{e}_{ij}$ that captures higher-order interactions between bonded atoms.

To ensure numerical stability and preserve the physically meaningful information encoded in the initial edge descriptors, a scaled residual update is applied:
\[
\mathbf{e}_{ij}^{\,\mathrm{new}} = \mathbf{e}_{ij} + 0.1\,\Delta \mathbf{e}_{ij}.
\]
This progressive refinement enables the model to adapt the edge features based on the evolving chemical and geometric context while avoiding abrupt feature distortions.

The updated edge features $\mathbf{e}_{ij}^{\,\mathrm{new}}$ are then directly used as inputs to the subsequent self-attention mechanism, ensuring that attention-based message passing is explicitly guided by the refined interatomic relationships.

\subsubsection{Edge-Aware Self-Attention via GATv2Conv}

Following the initial edge refinement step shown in Fig.~2(a), the updated edge features are immediately utilized within the edge-aware self-attention mechanism, illustrated in Fig.~2(b). This sequential coupling ensures that attention weights are computed using chemically and geometrically refined interatomic information.

To accurately model the complex spatial and chemical dependencies governing atomic displacements, self-attention is implemented using the GATv2Conv operator. Unlike the original GAT formulation, GATv2Conv computes attention scores after feature transformation, which eliminates static attention limitations and allows interaction strengths to dynamically adapt based on evolving atomic environments. This property is essential for relaxed structure prediction, where atomic movements are strongly influenced by subtle variations in local bonding geometry.

For a central atom $i$ and one of its neighboring atoms $j \in \mathcal{N}(i)$, the attention coefficient $\alpha_{ij}$ is computed from the transformed node and updated edge features as
\begin{equation}
\alpha_{ij} =
\frac{
\exp\!\left(
\sigma\!\left(
a^\top \left[ W x_i \, \| \, W x_j \, \| \, \mathbf{e}_{ij}^{\,\mathrm{new}} \right]
\right)
\right)
}{
\sum\limits_{k \in \mathcal{N}(i)}
\exp\!\left(
\sigma\!\left(
a^\top \left[ W x_i \, \| \, W x_k \, \| \, \mathbf{e}_{ik}^{\,\mathrm{new}} \right]
\right)
\right)
},
\end{equation}
where $W$ is a learnable linear transformation applied to node features, $a$ is the attention weight vector, $\|$ denotes concatenation, and $\sigma(\cdot)$ represents the LeakyReLU activation function.

This formulation allows each atom to dynamically evaluate the relative influence of its neighbors by jointly considering atomic identity, learned feature representation, and refined geometric–chemical bond information. Unlike fixed neighbor averaging, self-attention enables the model to assign unequal importance to different neighbors, which is physically meaningful because atomic displacements in relaxed structures are driven by uneven force contributions from surrounding atoms.

Once the attention coefficients are determined, the updated node feature of atom $i$ is obtained through attention-weighted aggregation:
\begin{equation}
h_i' = \sigma\left( \sum_{j \in \mathcal{N}(i)} \alpha_{ij} \, W x_j \right).
\end{equation}

Although Fig.~2(b) illustrates the attention process using a single head for clarity, the actual model employs a multi-head attention formulation to improve representational capacity. With $K$ independent heads, the node update becomes
\begin{equation}
h_i^{\prime (k)} =
\sigma\left( \sum_{j \in \mathcal{N}(i)} \alpha_{ij}^{(k)} \, W^{(k)} x_j \right), \quad k = 1,\dots,K,
\end{equation}
and the final embedding is obtained by concatenating all head outputs:
\begin{equation}
h_i' = \big\|_{k=1}^{K} h_i^{\prime (k)} .
\end{equation}

By integrating self-attention with dynamically refined edge features through GATv2Conv, the proposed EGAT framework enables direction-sensitive, chemistry-aware message passing. This is crucial for relaxed atomic position prediction, where atomic movements emerge from anisotropic and environment-dependent force contributions rather than uniform neighbor influence.

\subsubsection{Message Passing Block}

Following the computation of attention coefficients, the model performs edge-aware message passing as depicted in Fig.~2(c). In this stage, each neighboring atom $j$ communicates information to the central atom $i$ through a directed edge $(j,i)$ carrying the refined edge feature $\mathbf{e}_{ij}^{\,\text{new}}$.

For each atomic pair $(i,j)$, a message is generated using the concatenated node and edge representations,
\begin{equation}
\mathbf{m}_{ij} = \psi \left( [ \mathbf{x}_i \, \| \, \mathbf{x}_j \, \| \, \mathbf{e}_{ij}^{\,\text{new}} ] \right),
\end{equation}
where $\psi(\cdot)$ denotes a learnable neural transformation consistent with the attention and update operations.

Each message $\mathbf{m}_{ij}$ is then weighted by the corresponding attention coefficient $\alpha_{ij}$, ensuring that more influential neighbors contribute more strongly to the information flow. This attention-guided message passing allows the network to model anisotropic and chemically selective interatomic interactions in a physically meaningful way.

\subsubsection{Node Update Block}

As illustrated in Fig.~2(d), the attention-weighted messages from all neighbors are aggregated to update the feature representation of the central atom. For each atom $i$, the aggregated message is computed as
\begin{equation}
\mathbf{h}_i = \sum_{j \in \mathcal{N}(i)} \alpha_{ij} \, W \mathbf{x}_j,
\end{equation}
where $W$ is a learnable linear transformation applied to neighbor features.

To preserve previously learned information and stabilize training, a residual connection is applied by adding the original node feature $\mathbf{x}_i$ to the aggregated message. The resulting feature is then passed through batch normalization, ReLU activation, and dropout:
\begin{equation}
\mathbf{x}_i' =
\mathrm{Dropout}\!\left(
\mathrm{ReLU}\!\left(
\mathrm{BatchNorm}(\mathbf{h}_i + \mathbf{x}_i)
\right)
\right).
\end{equation}

This residual update prevents over-smoothing and ensures efficient gradient propagation across deep EGAT stacks. The updated node features $\mathbf{x}_i'$ are then forwarded to the next EGAT block, where the sequence of edge update, attention, message passing, and node update is repeated.

After passing through three successive EGAT blocks, the final node embeddings encode multi-hop chemical and geometric interactions. These embeddings are mapped through a fully connected output layer to predict the relaxed atomic positions,
\begin{equation}
\hat{\mathbf{p}}_i = W_{\mathrm{out}} \mathbf{x}_i', \quad \hat{\mathbf{p}}_i \in \mathbb{R}^3,
\end{equation}
where $\hat{\mathbf{p}}_i$ denotes the predicted relaxed position of atom $i$.
We evaluated the performance of the proposed edge-aware GAT model on a diverse set of high-entropy carbide (HEC) systems spanning binary to quinary compositions. The predictive capability of the model was systematically assessed using parity plots of atomic displacements, error distribution analysis, directional mean absolute errors, and predictive uncertainty estimation. Together, these metrics provide a comprehensive validation of the ability of model to reproduce relaxed structure positions of chemically complex systems.

\par
\begin{figure}[t]
    \centering
    \vspace{2mm}
    \includegraphics[width=1.0\linewidth,height=7cm,keepaspectratio]{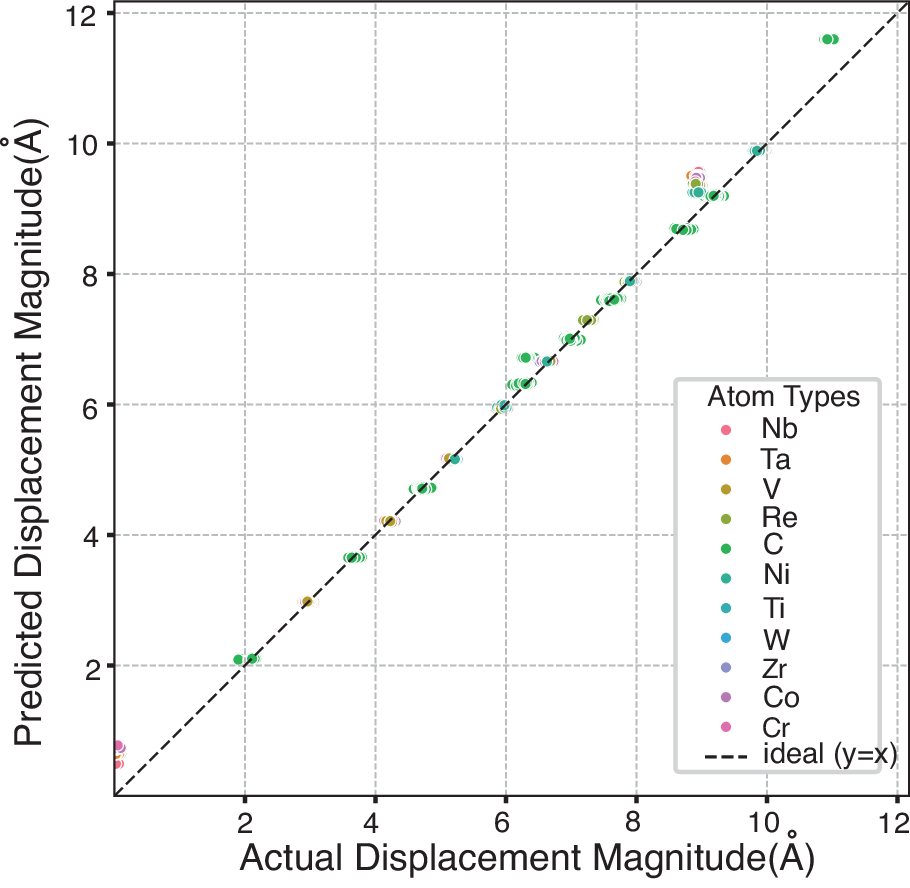}
    \caption{Parity plot of predicted versus actual atomic displacement magnitudes for the test set. The close alignment of data points along the $y=x$ line indicates high agreement with DFT-relaxed positions.}
    \label{fig3}
    \vspace{-2mm}
\end{figure}

 Figure~\ref{fig3} presents the parity plot between the predicted and DFT-computed atomic displacement magnitudes for the test set. The displacement magnitude is defined as the Euclidean norm $\lVert \mathbf{r}_{\text{final}} - \mathbf{r}_{\text{initial}} \rVert$, where $\mathbf{r}_{\text{initial}}$ and $\mathbf{r}_{\text{final}}$ denote the initial and relaxed atomic positions, respectively. The dashed diagonal line represents the ideal prediction ($y=x$). The strong clustering of data points tightly along this diagonal across the full displacement range (up to $\sim 11$~\AA) demonstrates the high predictive accuracy of the model. The achieved Mean Absolute Error (MAE) of $0.07$~\AA\ and Root Mean Square Error (RMSE) of $0.17$~\AA\ confirm that the model closely reproduces DFT-level structural relaxations.

\par The color-coded representation of different atomic species (\(\text{Nb, Ta, V, Re, C, Ni, Ti, W, Zr, Co, Cr}\)) further demonstrates the chemical generalization capability of the model. The uniform alignment of all atomic types along the diagonal indicates the absence of elemental bias and confirms that the model can accurately predict both light interstitial atoms and heavy metallic elements. The vertical clustering observed at higher displacements reflects the collective relaxation of atoms within the same crystal lattice, confirming that the model captures correlated structural responses consistently across multi-component systems.

\begin{figure}[t]
    \centering
    \vspace{2mm}
    \includegraphics[width=1.0\linewidth,height=7cm,keepaspectratio]{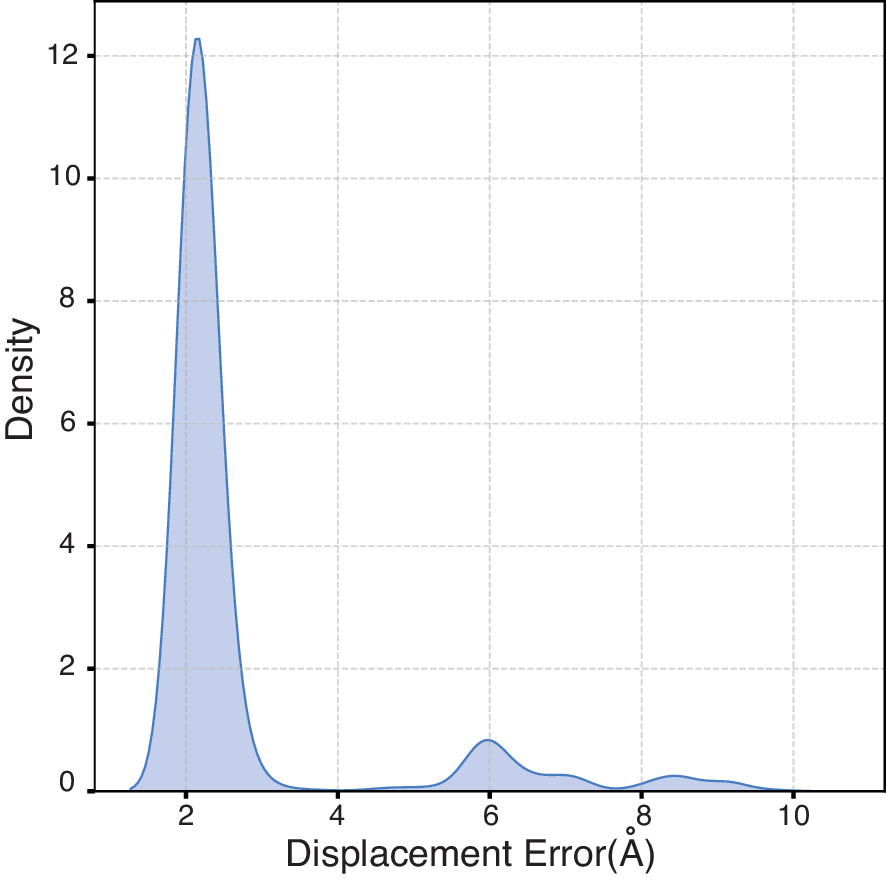}
    \caption{Probability density distribution of atomic displacement prediction errors for the test set.}
    \label{fig4}
    \vspace{-2mm}
\end{figure}

\par Figure~\ref{fig4} shows the probability density distribution of the absolute atomic displacement errors. A sharp and dominant peak at low error values indicates that the majority of atomic displacements are predicted with high precision. The narrow width of the main peak confirms that errors are tightly concentrated near zero, demonstrating reliable learning of local atomic environments. The small tail extending toward higher errors corresponds to a limited number of atoms in complex coordination environments or high-strain regions, where structural relaxation is inherently more difficult to predict. Overall, the narrow and right-skewed error distribution confirms the robustness and physical consistency of the model across diverse configurational environments.

\begin{figure}[t]
    \centering
    \vspace{2mm}
    \includegraphics[width=1.0\linewidth,height=7cm,keepaspectratio]{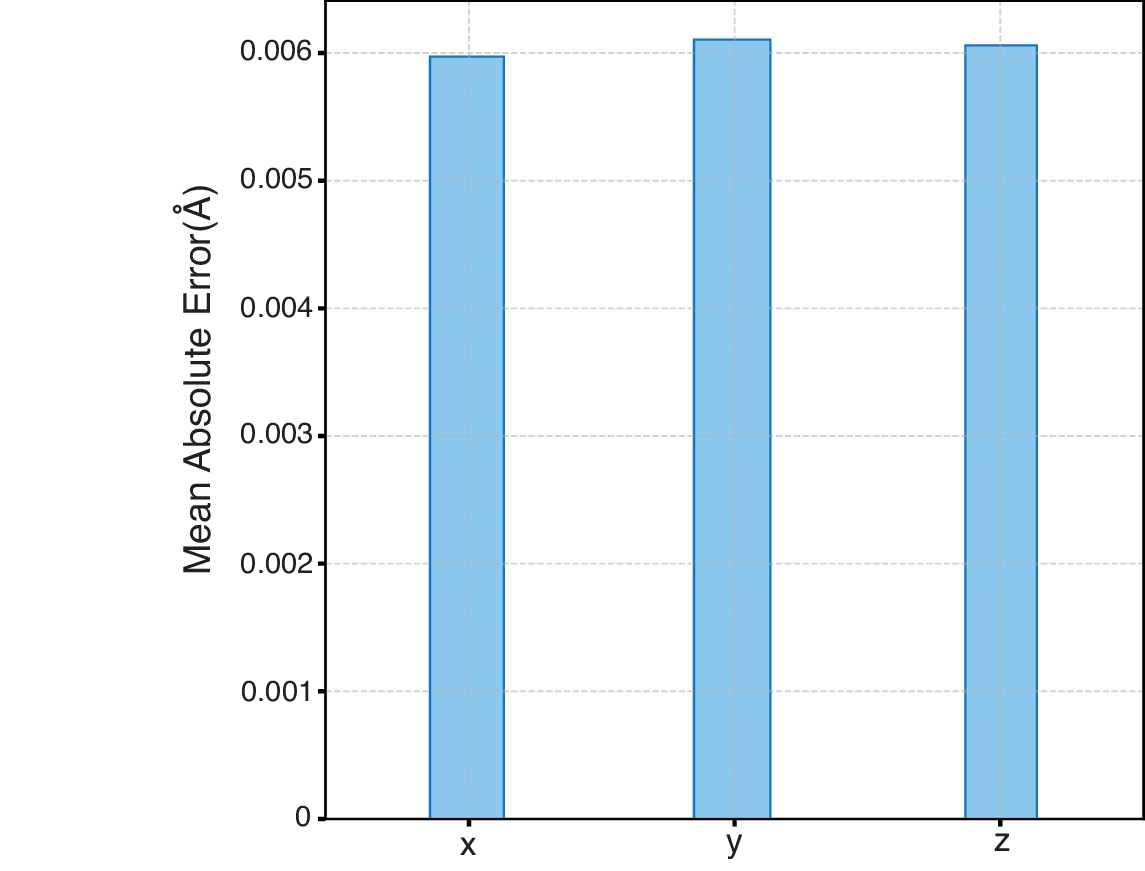}
    \caption{Mean absolute error (MAE) of predicted atomic positions along the $x$, $y$, and $z$ directions.}
    \label{fig5}
    \vspace{-2mm}
\end{figure}

\par Figure~\ref{fig5} reports the directional mean absolute errors along the $x$, $y$, and $z$ axes. The nearly identical MAE values in all three directions indicate that the model does not exhibit any directional bias and treats all spatial dimensions equivalently. This isotropic error behavior confirms that the model preserves the rotational symmetry of the physical system and accurately learns vector-valued atomic displacements in three-dimensional space. Such symmetry-consistent predictions are essential for reliable structure relaxation and subsequent property predictions.

\begin{figure}[t]
    \centering
    \vspace{2mm}
    \includegraphics[width=1.0\linewidth,height=7cm,keepaspectratio]{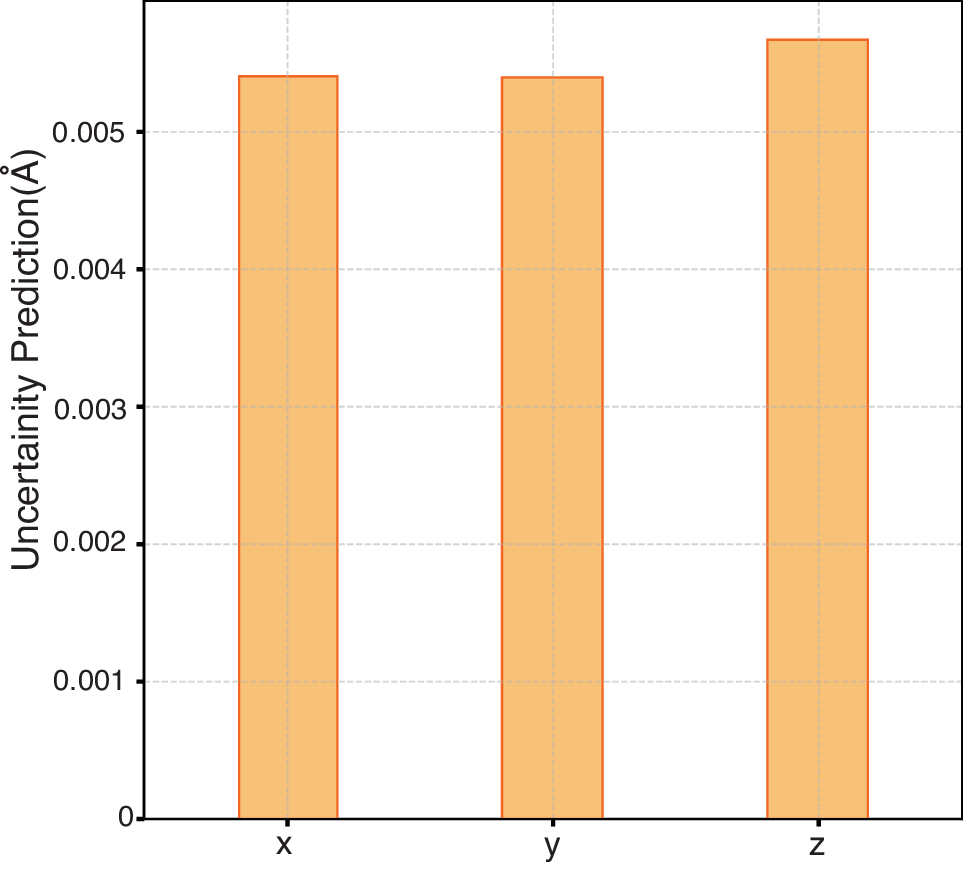}
    \caption{Predictive uncertainty of the relaxed atomic positions along the Cartesian directions ($x$, $y$, and $z$) estimated using Monte Carlo (MC) Dropout with 30 stochastic forward passes. The nearly identical and low uncertainty values across all three directions indicate isotropic model behaviour and high confidence in the predicted atomic structures.}
    \label{fig6}
    \vspace{-2mm}
\end{figure}

\subsubsection{Uncertainty Quantification Using MC Dropout}

In atomistic machine learning, predictive uncertainty plays a crucial role in assessing the reliability and robustness of model predictions, particularly for out-of-distribution structures and chemically complex systems such as high-entropy carbides. Unlike deterministic predictions, uncertainty-aware models provide confidence estimates alongside predicted atomic positions, which is essential for guiding high-throughput screening, active learning, and decision-making in materials discovery.

To quantify the epistemic uncertainty of our edge-aware GAT model, we employ Monte Carlo (MC) Dropout, a widely used approximation to Bayesian inference in deep neural networks. During inference, dropout layers are kept active and multiple stochastic forward passes are performed with different dropout masks. The variance in the predicted atomic positions across these stochastic samples is then used as a measure of predictive uncertainty. In this work, 30 stochastic forward passes were performed to estimate uncertainty for each atomic coordinate.

Figure~\ref{fig6} presents the predictive uncertainty along the Cartesian directions ($x$, $y$, and $z$). The uncertainty values are consistently low, remaining close to $\sim 0.005$--$0.006$~\AA~across all spatial directions. The nearly identical uncertainty in $x$, $y$, and $z$ confirms that the model treats all coordinate directions uniformly and does not exhibit directional bias. This isotropic uncertainty behavior is a direct consequence of the rotational equivariance built into the model architecture. Furthermore, the low magnitude and stability of the uncertainty indicate that the model produces confident and reliable predictions for relaxed atomic structures.

Quantitatively, the edge-aware GAT model achieves a mean absolute error (MAE) of 0.09 \AA{} and a root mean square error (RMSE) of 0.18 \AA{} on the test set. These results highlight the strength of the model in capturing complex atomic interactions using attention-based message passing and chemically informed descriptors. Together, the visual and numerical evidence demonstrate that the model can serve as an efficient and accurate surrogate for DFT relaxations, delivering near DFT-level precision at a fraction of the computational cost.
\section{Conclusion}
In this work, we introduced an edge-aware Graph Attention Network (EGAT) specifically developed for predicting relaxed atomic structures of high-entropy carbide systems. By integrating chemically informed descriptors inspired by the CLEAR framework with a combination of invariant and equivariant geometric features, the proposed model effectively captures local atomic environments, directional bonding information, and symmetry-aware structural relationships.

The framework employs edge-aware attention and iterative message passing to jointly update node and edge representations based on both atomic positions and chemical identity. As a result, the model learns a physically meaningful mapping from ideal to relaxed atomic configurations with high accuracy.

Compared to conventional density functional theory (DFT) calculations, the proposed model achieves accurate structural relaxation at a fraction of the computational cost, enabling rapid large-scale screening of chemically complex carbide systems and efficient exploration of high-dimensional compositional spaces. 

Beyond structural relaxation, this framework establishes a general, extensible methodology that can be directly applied to predict formation energies, mechanical properties, and other structure-dependent physical quantities. By providing fast and reliable access to relaxed geometries, the model serves as a foundational tool for future data-driven studies in materials stability, phase behavior, and functional property optimization. Overall, the edge-aware GAT offers a scalable and cost-effective platform for accelerating scientific discovery and experimental design in high-entropy materials research.

\section{Data availability}
The database and codes used for this study are openly available in GitHub.
\section{CRediT authorship contribution statement}


\section{Acknowledgements}
The authors thank the Materials Research Centre (MRC) and Solid State and Structural Chemistry Unit (SSCU), Indian Institute of Science, for the computational facilities. Neethu Mohan Mangalassery acknowledges the UGC Research fellowship. The authors acknowledge the support from the Institute of Eminence (IoE) scheme of The Ministry of Human Resource Development, Government of India.

\bibliography{manuscript}

\providecommand{\noopsort}[1]{}\providecommand{\singleletter}[1]{#1}%
\begin{thebibliography}{33}%
\makeatletter
\providecommand \@ifxundefined [1]{%
 \@ifx{#1\undefined}
}%
\providecommand \@ifnum [1]{%
 \ifnum #1\expandafter \@firstoftwo
 \else \expandafter \@secondoftwo
 \fi
}%
\providecommand \@ifx [1]{%
 \ifx #1\expandafter \@firstoftwo
 \else \expandafter \@secondoftwo
 \fi
}%
\providecommand \natexlab [1]{#1}%
\providecommand \enquote  [1]{``#1''}%
\providecommand \bibnamefont  [1]{#1}%
\providecommand \bibfnamefont [1]{#1}%
\providecommand \citenamefont [1]{#1}%
\providecommand \href@noop [0]{\@secondoftwo}%
\providecommand \href [0]{\begingroup \@sanitize@url \@href}%
\providecommand \@href[1]{\@@startlink{#1}\@@href}%
\providecommand \@@href[1]{\endgroup#1\@@endlink}%
\providecommand \@sanitize@url [0]{\catcode `\\12\catcode `\$12\catcode
  `\&12\catcode `\#12\catcode `\^12\catcode `\_12\catcode `\%12\relax}%
\providecommand \@@startlink[1]{}%
\providecommand \@@endlink[0]{}%
\providecommand \url  [0]{\begingroup\@sanitize@url \@url }%
\providecommand \@url [1]{\endgroup\@href {#1}{\urlprefix }}%
\providecommand \urlprefix  [0]{URL }%
\providecommand \Eprint [0]{\href }%
\providecommand \doibase [0]{https://doi.org/}%
\providecommand \selectlanguage [0]{\@gobble}%
\providecommand \bibinfo  [0]{\@secondoftwo}%
\providecommand \bibfield  [0]{\@secondoftwo}%
\providecommand \translation [1]{[#1]}%
\providecommand \BibitemOpen [0]{}%
\providecommand \bibitemStop [0]{}%
\providecommand \bibitemNoStop [0]{.\EOS\space}%
\providecommand \EOS [0]{\spacefactor3000\relax}%
\providecommand \BibitemShut  [1]{\csname bibitem#1\endcsname}%
\let\auto@bib@innerbib\@empty
\bibitem [{\citenamefont {Schweidler}\ \emph {et~al.}(2024)\citenamefont
  {Schweidler}, \citenamefont {Botros}, \citenamefont {Strauss}, \citenamefont
  {Wang}, \citenamefont {Ma}, \citenamefont {Velasco}, \citenamefont
  {Cadilha~Marques}, \citenamefont {Sarkar}, \citenamefont {K{\"u}bel},
  \citenamefont {Hahn} \emph {et~al.}}]{schweidler2024high}%
  \BibitemOpen
  \bibfield  {author} {\bibinfo {author} {\bibfnamefont {S.}~\bibnamefont
  {Schweidler}}, \bibinfo {author} {\bibfnamefont {M.}~\bibnamefont {Botros}},
  \bibinfo {author} {\bibfnamefont {F.}~\bibnamefont {Strauss}}, \bibinfo
  {author} {\bibfnamefont {Q.}~\bibnamefont {Wang}}, \bibinfo {author}
  {\bibfnamefont {Y.}~\bibnamefont {Ma}}, \bibinfo {author} {\bibfnamefont
  {L.}~\bibnamefont {Velasco}}, \bibinfo {author} {\bibfnamefont
  {G.}~\bibnamefont {Cadilha~Marques}}, \bibinfo {author} {\bibfnamefont
  {A.}~\bibnamefont {Sarkar}}, \bibinfo {author} {\bibfnamefont
  {C.}~\bibnamefont {K{\"u}bel}}, \bibinfo {author} {\bibfnamefont
  {H.}~\bibnamefont {Hahn}}, \emph {et~al.},\ }\href@noop {} {\bibfield
  {journal} {\bibinfo  {journal} {Nature Reviews Materials}\ }\textbf {\bibinfo
  {volume} {9}},\ \bibinfo {pages} {266} (\bibinfo {year} {2024})}\BibitemShut
  {NoStop}%
\bibitem [{\citenamefont {Ren}\ \emph {et~al.}(2025)\citenamefont {Ren},
  \citenamefont {Kumkale}, \citenamefont {Hou}, \citenamefont {Kadam},
  \citenamefont {Jagtap}, \citenamefont {Lokhande}, \citenamefont {Pathan},
  \citenamefont {Pereira}, \citenamefont {Lei},\ and\ \citenamefont
  {Liu}}]{ren2025review}%
  \BibitemOpen
  \bibfield  {author} {\bibinfo {author} {\bibfnamefont {J.}~\bibnamefont
  {Ren}}, \bibinfo {author} {\bibfnamefont {V.~Y.}\ \bibnamefont {Kumkale}},
  \bibinfo {author} {\bibfnamefont {H.}~\bibnamefont {Hou}}, \bibinfo {author}
  {\bibfnamefont {V.~S.}\ \bibnamefont {Kadam}}, \bibinfo {author}
  {\bibfnamefont {C.~V.}\ \bibnamefont {Jagtap}}, \bibinfo {author}
  {\bibfnamefont {P.~E.}\ \bibnamefont {Lokhande}}, \bibinfo {author}
  {\bibfnamefont {H.~M.}\ \bibnamefont {Pathan}}, \bibinfo {author}
  {\bibfnamefont {A.}~\bibnamefont {Pereira}}, \bibinfo {author} {\bibfnamefont
  {H.}~\bibnamefont {Lei}},\ and\ \bibinfo {author} {\bibfnamefont {T.~X.}\
  \bibnamefont {Liu}},\ }\href@noop {} {\bibfield  {journal} {\bibinfo
  {journal} {Advanced Composites and Hybrid Materials}\ }\textbf {\bibinfo
  {volume} {8}},\ \bibinfo {pages} {195} (\bibinfo {year} {2025})}\BibitemShut
  {NoStop}%
\bibitem [{\citenamefont {Praveen}\ and\ \citenamefont
  {Kim}(2018)}]{praveen2018high}%
  \BibitemOpen
  \bibfield  {author} {\bibinfo {author} {\bibfnamefont {S.}~\bibnamefont
  {Praveen}}\ and\ \bibinfo {author} {\bibfnamefont {H.~S.}\ \bibnamefont
  {Kim}},\ }\href@noop {} {\bibfield  {journal} {\bibinfo  {journal} {Advanced
  Engineering Materials}\ }\textbf {\bibinfo {volume} {20}},\ \bibinfo {pages}
  {1700645} (\bibinfo {year} {2018})}\BibitemShut {NoStop}%
\bibitem [{\citenamefont {Wang}\ \emph {et~al.}(2021)\citenamefont {Wang},
  \citenamefont {Guo},\ and\ \citenamefont {Fu}}]{wang2021high}%
  \BibitemOpen
  \bibfield  {author} {\bibinfo {author} {\bibfnamefont {X.}~\bibnamefont
  {Wang}}, \bibinfo {author} {\bibfnamefont {W.}~\bibnamefont {Guo}},\ and\
  \bibinfo {author} {\bibfnamefont {Y.}~\bibnamefont {Fu}},\ }\href@noop {}
  {\bibfield  {journal} {\bibinfo  {journal} {Journal of Materials Chemistry
  A}\ }\textbf {\bibinfo {volume} {9}},\ \bibinfo {pages} {663} (\bibinfo
  {year} {2021})}\BibitemShut {NoStop}%
\bibitem [{\citenamefont {Miracle}\ \emph {et~al.}(2014)\citenamefont
  {Miracle}, \citenamefont {Miller}, \citenamefont {Senkov}, \citenamefont
  {Woodward}, \citenamefont {Uchic},\ and\ \citenamefont
  {Tiley}}]{miracle2014exploration}%
  \BibitemOpen
  \bibfield  {author} {\bibinfo {author} {\bibfnamefont {D.~B.}\ \bibnamefont
  {Miracle}}, \bibinfo {author} {\bibfnamefont {J.~D.}\ \bibnamefont {Miller}},
  \bibinfo {author} {\bibfnamefont {O.~N.}\ \bibnamefont {Senkov}}, \bibinfo
  {author} {\bibfnamefont {C.}~\bibnamefont {Woodward}}, \bibinfo {author}
  {\bibfnamefont {M.~D.}\ \bibnamefont {Uchic}},\ and\ \bibinfo {author}
  {\bibfnamefont {J.}~\bibnamefont {Tiley}},\ }\href@noop {} {\bibfield
  {journal} {\bibinfo  {journal} {Entropy}\ }\textbf {\bibinfo {volume} {16}},\
  \bibinfo {pages} {494} (\bibinfo {year} {2014})}\BibitemShut {NoStop}%
\bibitem [{\citenamefont {Fan}\ \emph {et~al.}(2024)\citenamefont {Fan},
  \citenamefont {Sun}, \citenamefont {Zhao},\ and\ \citenamefont
  {Yun}}]{fan2024high}%
  \BibitemOpen
  \bibfield  {author} {\bibinfo {author} {\bibfnamefont {C.}~\bibnamefont
  {Fan}}, \bibinfo {author} {\bibfnamefont {J.}~\bibnamefont {Sun}}, \bibinfo
  {author} {\bibfnamefont {J.}~\bibnamefont {Zhao}},\ and\ \bibinfo {author}
  {\bibfnamefont {X.}~\bibnamefont {Yun}},\ }\href@noop {} {\bibfield
  {journal} {\bibinfo  {journal} {Ceramics International}\ } (\bibinfo {year}
  {2024})}\BibitemShut {NoStop}%
\bibitem [{\citenamefont {Sarker}\ \emph {et~al.}(2018)\citenamefont {Sarker},
  \citenamefont {Harrington}, \citenamefont {Toher}, \citenamefont {Oses},
  \citenamefont {Samiee}, \citenamefont {Maria}, \citenamefont {Brenner},
  \citenamefont {Vecchio},\ and\ \citenamefont {Curtarolo}}]{sarker2018high}%
  \BibitemOpen
  \bibfield  {author} {\bibinfo {author} {\bibfnamefont {P.}~\bibnamefont
  {Sarker}}, \bibinfo {author} {\bibfnamefont {T.}~\bibnamefont {Harrington}},
  \bibinfo {author} {\bibfnamefont {C.}~\bibnamefont {Toher}}, \bibinfo
  {author} {\bibfnamefont {C.}~\bibnamefont {Oses}}, \bibinfo {author}
  {\bibfnamefont {M.}~\bibnamefont {Samiee}}, \bibinfo {author} {\bibfnamefont
  {J.-P.}\ \bibnamefont {Maria}}, \bibinfo {author} {\bibfnamefont {D.~W.}\
  \bibnamefont {Brenner}}, \bibinfo {author} {\bibfnamefont {K.~S.}\
  \bibnamefont {Vecchio}},\ and\ \bibinfo {author} {\bibfnamefont
  {S.}~\bibnamefont {Curtarolo}},\ }\href@noop {} {\bibfield  {journal}
  {\bibinfo  {journal} {Nature communications}\ }\textbf {\bibinfo {volume}
  {9}},\ \bibinfo {pages} {4980} (\bibinfo {year} {2018})}\BibitemShut
  {NoStop}%
\bibitem [{\citenamefont {Feltrin}\ \emph {et~al.}(2022)\citenamefont
  {Feltrin}, \citenamefont {Xing}, \citenamefont {Akinwekomi}, \citenamefont
  {Waseem},\ and\ \citenamefont {Akhtar}}]{feltrin2022review}%
  \BibitemOpen
  \bibfield  {author} {\bibinfo {author} {\bibfnamefont {A.~C.}\ \bibnamefont
  {Feltrin}}, \bibinfo {author} {\bibfnamefont {Q.}~\bibnamefont {Xing}},
  \bibinfo {author} {\bibfnamefont {A.~D.}\ \bibnamefont {Akinwekomi}},
  \bibinfo {author} {\bibfnamefont {O.~A.}\ \bibnamefont {Waseem}},\ and\
  \bibinfo {author} {\bibfnamefont {F.}~\bibnamefont {Akhtar}},\ }\href@noop {}
  {\bibfield  {journal} {\bibinfo  {journal} {Entropy}\ }\textbf {\bibinfo
  {volume} {25}},\ \bibinfo {pages} {73} (\bibinfo {year} {2022})}\BibitemShut
  {NoStop}%
\bibitem [{\citenamefont {Orio}\ \emph {et~al.}(2009)\citenamefont {Orio},
  \citenamefont {Pantazis},\ and\ \citenamefont {Neese}}]{orio2009density}%
  \BibitemOpen
  \bibfield  {author} {\bibinfo {author} {\bibfnamefont {M.}~\bibnamefont
  {Orio}}, \bibinfo {author} {\bibfnamefont {D.~A.}\ \bibnamefont {Pantazis}},\
  and\ \bibinfo {author} {\bibfnamefont {F.}~\bibnamefont {Neese}},\
  }\href@noop {} {\bibfield  {journal} {\bibinfo  {journal} {Photosynthesis
  research}\ }\textbf {\bibinfo {volume} {102}},\ \bibinfo {pages} {443}
  (\bibinfo {year} {2009})}\BibitemShut {NoStop}%
\bibitem [{\citenamefont {Geerlings}\ \emph {et~al.}(2003)\citenamefont
  {Geerlings}, \citenamefont {De~Proft},\ and\ \citenamefont
  {Langenaeker}}]{geerlings2003conceptual}%
  \BibitemOpen
  \bibfield  {author} {\bibinfo {author} {\bibfnamefont {P.}~\bibnamefont
  {Geerlings}}, \bibinfo {author} {\bibfnamefont {F.}~\bibnamefont
  {De~Proft}},\ and\ \bibinfo {author} {\bibfnamefont {W.}~\bibnamefont
  {Langenaeker}},\ }\href@noop {} {\bibfield  {journal} {\bibinfo  {journal}
  {Chemical reviews}\ }\textbf {\bibinfo {volume} {103}},\ \bibinfo {pages}
  {1793} (\bibinfo {year} {2003})}\BibitemShut {NoStop}%
\bibitem [{\citenamefont {Mishin}(2021)}]{mishin2021machine}%
  \BibitemOpen
  \bibfield  {author} {\bibinfo {author} {\bibfnamefont {Y.}~\bibnamefont
  {Mishin}},\ }\href@noop {} {\bibfield  {journal} {\bibinfo  {journal} {Acta
  Materialia}\ }\textbf {\bibinfo {volume} {214}},\ \bibinfo {pages} {116980}
  (\bibinfo {year} {2021})}\BibitemShut {NoStop}%
\bibitem [{\citenamefont {Kulichenko}\ \emph {et~al.}(2024)\citenamefont
  {Kulichenko}, \citenamefont {Nebgen}, \citenamefont {Lubbers}, \citenamefont
  {Smith}, \citenamefont {Barros}, \citenamefont {Allen}, \citenamefont
  {Habib}, \citenamefont {Shinkle}, \citenamefont {Fedik}, \citenamefont {Li}
  \emph {et~al.}}]{kulichenko2024data}%
  \BibitemOpen
  \bibfield  {author} {\bibinfo {author} {\bibfnamefont {M.}~\bibnamefont
  {Kulichenko}}, \bibinfo {author} {\bibfnamefont {B.}~\bibnamefont {Nebgen}},
  \bibinfo {author} {\bibfnamefont {N.}~\bibnamefont {Lubbers}}, \bibinfo
  {author} {\bibfnamefont {J.~S.}\ \bibnamefont {Smith}}, \bibinfo {author}
  {\bibfnamefont {K.}~\bibnamefont {Barros}}, \bibinfo {author} {\bibfnamefont
  {A.~E.}\ \bibnamefont {Allen}}, \bibinfo {author} {\bibfnamefont
  {A.}~\bibnamefont {Habib}}, \bibinfo {author} {\bibfnamefont
  {E.}~\bibnamefont {Shinkle}}, \bibinfo {author} {\bibfnamefont
  {N.}~\bibnamefont {Fedik}}, \bibinfo {author} {\bibfnamefont {Y.~W.}\
  \bibnamefont {Li}}, \emph {et~al.},\ }\href@noop {} {\bibfield  {journal}
  {\bibinfo  {journal} {Chemical Reviews}\ }\textbf {\bibinfo {volume} {124}},\
  \bibinfo {pages} {13681} (\bibinfo {year} {2024})}\BibitemShut {NoStop}%
\bibitem [{\citenamefont {Novikov}\ \emph {et~al.}(2020)\citenamefont
  {Novikov}, \citenamefont {Gubaev}, \citenamefont {Podryabinkin},\ and\
  \citenamefont {Shapeev}}]{novikov2020mlip}%
  \BibitemOpen
  \bibfield  {author} {\bibinfo {author} {\bibfnamefont {I.~S.}\ \bibnamefont
  {Novikov}}, \bibinfo {author} {\bibfnamefont {K.}~\bibnamefont {Gubaev}},
  \bibinfo {author} {\bibfnamefont {E.~V.}\ \bibnamefont {Podryabinkin}},\ and\
  \bibinfo {author} {\bibfnamefont {A.~V.}\ \bibnamefont {Shapeev}},\
  }\href@noop {} {\bibfield  {journal} {\bibinfo  {journal} {Machine Learning:
  Science and Technology}\ }\textbf {\bibinfo {volume} {2}},\ \bibinfo {pages}
  {025002} (\bibinfo {year} {2020})}\BibitemShut {NoStop}%
\bibitem [{\citenamefont {Yang}\ \emph {et~al.}(2024)\citenamefont {Yang},
  \citenamefont {Zhao}, \citenamefont {Wang}, \citenamefont {Liu},
  \citenamefont {Zhang}, \citenamefont {Li}, \citenamefont {Lv}, \citenamefont
  {Chen},\ and\ \citenamefont {Shen}}]{yang2024scalable}%
  \BibitemOpen
  \bibfield  {author} {\bibinfo {author} {\bibfnamefont {Z.}~\bibnamefont
  {Yang}}, \bibinfo {author} {\bibfnamefont {Y.-M.}\ \bibnamefont {Zhao}},
  \bibinfo {author} {\bibfnamefont {X.}~\bibnamefont {Wang}}, \bibinfo {author}
  {\bibfnamefont {X.}~\bibnamefont {Liu}}, \bibinfo {author} {\bibfnamefont
  {X.}~\bibnamefont {Zhang}}, \bibinfo {author} {\bibfnamefont
  {Y.}~\bibnamefont {Li}}, \bibinfo {author} {\bibfnamefont {Q.}~\bibnamefont
  {Lv}}, \bibinfo {author} {\bibfnamefont {C.~Y.-C.}\ \bibnamefont {Chen}},\
  and\ \bibinfo {author} {\bibfnamefont {L.}~\bibnamefont {Shen}},\ }\href@noop
  {} {\bibfield  {journal} {\bibinfo  {journal} {Nature Communications}\
  }\textbf {\bibinfo {volume} {15}},\ \bibinfo {pages} {8148} (\bibinfo {year}
  {2024})}\BibitemShut {NoStop}%
\bibitem [{\citenamefont {Gibson}\ \emph {et~al.}(2022)\citenamefont {Gibson},
  \citenamefont {Hire},\ and\ \citenamefont {Hennig}}]{gibson2022data}%
  \BibitemOpen
  \bibfield  {author} {\bibinfo {author} {\bibfnamefont {J.}~\bibnamefont
  {Gibson}}, \bibinfo {author} {\bibfnamefont {A.}~\bibnamefont {Hire}},\ and\
  \bibinfo {author} {\bibfnamefont {R.~G.}\ \bibnamefont {Hennig}},\
  }\href@noop {} {\bibfield  {journal} {\bibinfo  {journal} {npj Computational
  Materials}\ }\textbf {\bibinfo {volume} {8}},\ \bibinfo {pages} {211}
  (\bibinfo {year} {2022})}\BibitemShut {NoStop}%
\bibitem [{\citenamefont {Kim}\ \emph {et~al.}(2023)\citenamefont {Kim},
  \citenamefont {Noh}, \citenamefont {Jin}, \citenamefont {Lee},\ and\
  \citenamefont {Jung}}]{kim2023structure}%
  \BibitemOpen
  \bibfield  {author} {\bibinfo {author} {\bibfnamefont {S.}~\bibnamefont
  {Kim}}, \bibinfo {author} {\bibfnamefont {J.}~\bibnamefont {Noh}}, \bibinfo
  {author} {\bibfnamefont {T.}~\bibnamefont {Jin}}, \bibinfo {author}
  {\bibfnamefont {J.}~\bibnamefont {Lee}},\ and\ \bibinfo {author}
  {\bibfnamefont {Y.}~\bibnamefont {Jung}},\ }\href@noop {} {\bibfield
  {journal} {\bibinfo  {journal} {npj Computational Materials}\ }\textbf
  {\bibinfo {volume} {9}},\ \bibinfo {pages} {142} (\bibinfo {year}
  {2023})}\BibitemShut {NoStop}%
\bibitem [{\citenamefont {Swetlana}\ and\ \citenamefont
  {Singh}(2024)}]{swetlana2024chemistry}%
  \BibitemOpen
  \bibfield  {author} {\bibinfo {author} {\bibfnamefont {S.}~\bibnamefont
  {Swetlana}}\ and\ \bibinfo {author} {\bibfnamefont {A.~K.}\ \bibnamefont
  {Singh}},\ }\href@noop {} {\bibfield  {journal} {\bibinfo  {journal} {Acta
  Materialia}\ ,\ \bibinfo {pages} {120122}} (\bibinfo {year}
  {2024})}\BibitemShut {NoStop}%
\bibitem [{\citenamefont {Zunger}\ \emph {et~al.}(1990)\citenamefont {Zunger},
  \citenamefont {Wei}, \citenamefont {Ferreira},\ and\ \citenamefont
  {Bernard}}]{zunger1990special}%
  \BibitemOpen
  \bibfield  {author} {\bibinfo {author} {\bibfnamefont {A.}~\bibnamefont
  {Zunger}}, \bibinfo {author} {\bibfnamefont {S.-H.}\ \bibnamefont {Wei}},
  \bibinfo {author} {\bibfnamefont {L.}~\bibnamefont {Ferreira}},\ and\
  \bibinfo {author} {\bibfnamefont {J.~E.}\ \bibnamefont {Bernard}},\
  }\href@noop {} {\bibfield  {journal} {\bibinfo  {journal} {Physical review
  letters}\ }\textbf {\bibinfo {volume} {65}},\ \bibinfo {pages} {353}
  (\bibinfo {year} {1990})}\BibitemShut {NoStop}%
\bibitem [{\citenamefont {Kresse}\ and\ \citenamefont
  {Furthm{\"u}ller}(1996{\natexlab{a}})}]{kresse1996efficient}%
  \BibitemOpen
  \bibfield  {author} {\bibinfo {author} {\bibfnamefont {G.}~\bibnamefont
  {Kresse}}\ and\ \bibinfo {author} {\bibfnamefont {J.}~\bibnamefont
  {Furthm{\"u}ller}},\ }\href@noop {} {\bibfield  {journal} {\bibinfo
  {journal} {Physical review B}\ }\textbf {\bibinfo {volume} {54}},\ \bibinfo
  {pages} {11169} (\bibinfo {year} {1996}{\natexlab{a}})}\BibitemShut {NoStop}%
\bibitem [{\citenamefont {Kresse}\ and\ \citenamefont
  {Furthm{\"u}ller}(1996{\natexlab{b}})}]{kresse1996efficiency}%
  \BibitemOpen
  \bibfield  {author} {\bibinfo {author} {\bibfnamefont {G.}~\bibnamefont
  {Kresse}}\ and\ \bibinfo {author} {\bibfnamefont {J.}~\bibnamefont
  {Furthm{\"u}ller}},\ }\href@noop {} {\bibfield  {journal} {\bibinfo
  {journal} {Computational materials science}\ }\textbf {\bibinfo {volume}
  {6}},\ \bibinfo {pages} {15} (\bibinfo {year}
  {1996}{\natexlab{b}})}\BibitemShut {NoStop}%
\bibitem [{\citenamefont {Bl{\"o}chl}(1994)}]{blochl1994projector}%
  \BibitemOpen
  \bibfield  {author} {\bibinfo {author} {\bibfnamefont {P.~E.}\ \bibnamefont
  {Bl{\"o}chl}},\ }\href@noop {} {\bibfield  {journal} {\bibinfo  {journal}
  {Physical review B}\ }\textbf {\bibinfo {volume} {50}},\ \bibinfo {pages}
  {17953} (\bibinfo {year} {1994})}\BibitemShut {NoStop}%
\bibitem [{\citenamefont {Perdew}\ \emph {et~al.}(1996)\citenamefont {Perdew},
  \citenamefont {Burke},\ and\ \citenamefont
  {Ernzerhof}}]{perdew1996generalized}%
  \BibitemOpen
  \bibfield  {author} {\bibinfo {author} {\bibfnamefont {J.~P.}\ \bibnamefont
  {Perdew}}, \bibinfo {author} {\bibfnamefont {K.}~\bibnamefont {Burke}},\ and\
  \bibinfo {author} {\bibfnamefont {M.}~\bibnamefont {Ernzerhof}},\ }\href@noop
  {} {\bibfield  {journal} {\bibinfo  {journal} {Physical review letters}\
  }\textbf {\bibinfo {volume} {77}},\ \bibinfo {pages} {3865} (\bibinfo {year}
  {1996})}\BibitemShut {NoStop}%
\bibitem [{\citenamefont {Methfessel}\ and\ \citenamefont
  {Paxton}(1989)}]{methfessel1989high}%
  \BibitemOpen
  \bibfield  {author} {\bibinfo {author} {\bibfnamefont {M.}~\bibnamefont
  {Methfessel}}\ and\ \bibinfo {author} {\bibfnamefont {A.}~\bibnamefont
  {Paxton}},\ }\href@noop {} {\bibfield  {journal} {\bibinfo  {journal}
  {physical review B}\ }\textbf {\bibinfo {volume} {40}},\ \bibinfo {pages}
  {3616} (\bibinfo {year} {1989})}\BibitemShut {NoStop}%
\bibitem [{\citenamefont {Corso}\ \emph {et~al.}(2024)\citenamefont {Corso},
  \citenamefont {Stark}, \citenamefont {Jegelka}, \citenamefont {Jaakkola},\
  and\ \citenamefont {Barzilay}}]{corso2024graph}%
  \BibitemOpen
  \bibfield  {author} {\bibinfo {author} {\bibfnamefont {G.}~\bibnamefont
  {Corso}}, \bibinfo {author} {\bibfnamefont {H.}~\bibnamefont {Stark}},
  \bibinfo {author} {\bibfnamefont {S.}~\bibnamefont {Jegelka}}, \bibinfo
  {author} {\bibfnamefont {T.}~\bibnamefont {Jaakkola}},\ and\ \bibinfo
  {author} {\bibfnamefont {R.}~\bibnamefont {Barzilay}},\ }\href@noop {}
  {\bibfield  {journal} {\bibinfo  {journal} {Nature Reviews Methods Primers}\
  }\textbf {\bibinfo {volume} {4}},\ \bibinfo {pages} {17} (\bibinfo {year}
  {2024})}\BibitemShut {NoStop}%
\bibitem [{\citenamefont {Xie}\ and\ \citenamefont
  {Grossman}(2018)}]{xie2018crystal}%
  \BibitemOpen
  \bibfield  {author} {\bibinfo {author} {\bibfnamefont {T.}~\bibnamefont
  {Xie}}\ and\ \bibinfo {author} {\bibfnamefont {J.~C.}\ \bibnamefont
  {Grossman}},\ }\href@noop {} {\bibfield  {journal} {\bibinfo  {journal}
  {Physical review letters}\ }\textbf {\bibinfo {volume} {120}},\ \bibinfo
  {pages} {145301} (\bibinfo {year} {2018})}\BibitemShut {NoStop}%
\bibitem [{\citenamefont {Reiser}\ \emph {et~al.}(2022)\citenamefont {Reiser},
  \citenamefont {Neubert}, \citenamefont {Eberhard}, \citenamefont {Torresi},
  \citenamefont {Zhou}, \citenamefont {Shao}, \citenamefont {Metni},
  \citenamefont {van Hoesel}, \citenamefont {Schopmans}, \citenamefont {Sommer}
  \emph {et~al.}}]{reiser2022graph}%
  \BibitemOpen
  \bibfield  {author} {\bibinfo {author} {\bibfnamefont {P.}~\bibnamefont
  {Reiser}}, \bibinfo {author} {\bibfnamefont {M.}~\bibnamefont {Neubert}},
  \bibinfo {author} {\bibfnamefont {A.}~\bibnamefont {Eberhard}}, \bibinfo
  {author} {\bibfnamefont {L.}~\bibnamefont {Torresi}}, \bibinfo {author}
  {\bibfnamefont {C.}~\bibnamefont {Zhou}}, \bibinfo {author} {\bibfnamefont
  {C.}~\bibnamefont {Shao}}, \bibinfo {author} {\bibfnamefont {H.}~\bibnamefont
  {Metni}}, \bibinfo {author} {\bibfnamefont {C.}~\bibnamefont {van Hoesel}},
  \bibinfo {author} {\bibfnamefont {H.}~\bibnamefont {Schopmans}}, \bibinfo
  {author} {\bibfnamefont {T.}~\bibnamefont {Sommer}}, \emph {et~al.},\
  }\href@noop {} {\bibfield  {journal} {\bibinfo  {journal} {Communications
  Materials}\ }\textbf {\bibinfo {volume} {3}},\ \bibinfo {pages} {93}
  (\bibinfo {year} {2022})}\BibitemShut {NoStop}%
\bibitem [{\citenamefont {Korolev}\ \emph {et~al.}(2019)\citenamefont
  {Korolev}, \citenamefont {Mitrofanov}, \citenamefont {Korotcov},\ and\
  \citenamefont {Tkachenko}}]{korolev2019graph}%
  \BibitemOpen
  \bibfield  {author} {\bibinfo {author} {\bibfnamefont {V.}~\bibnamefont
  {Korolev}}, \bibinfo {author} {\bibfnamefont {A.}~\bibnamefont {Mitrofanov}},
  \bibinfo {author} {\bibfnamefont {A.}~\bibnamefont {Korotcov}},\ and\
  \bibinfo {author} {\bibfnamefont {V.}~\bibnamefont {Tkachenko}},\ }\href@noop
  {} {\bibfield  {journal} {\bibinfo  {journal} {Journal of chemical
  information and modeling}\ }\textbf {\bibinfo {volume} {60}},\ \bibinfo
  {pages} {22} (\bibinfo {year} {2019})}\BibitemShut {NoStop}%
\bibitem [{\citenamefont {Vrahatis}\ \emph {et~al.}(2024)\citenamefont
  {Vrahatis}, \citenamefont {Lazaros},\ and\ \citenamefont
  {Kotsiantis}}]{vrahatis2024graph}%
  \BibitemOpen
  \bibfield  {author} {\bibinfo {author} {\bibfnamefont {A.~G.}\ \bibnamefont
  {Vrahatis}}, \bibinfo {author} {\bibfnamefont {K.}~\bibnamefont {Lazaros}},\
  and\ \bibinfo {author} {\bibfnamefont {S.}~\bibnamefont {Kotsiantis}},\
  }\href@noop {} {\bibfield  {journal} {\bibinfo  {journal} {Future Internet}\
  }\textbf {\bibinfo {volume} {16}},\ \bibinfo {pages} {318} (\bibinfo {year}
  {2024})}\BibitemShut {NoStop}%
\bibitem [{\citenamefont {Xie}\ \emph {et~al.}(2020)\citenamefont {Xie},
  \citenamefont {Zhang}, \citenamefont {Gong}, \citenamefont {Tang},\ and\
  \citenamefont {Han}}]{xie2020mgat}%
  \BibitemOpen
  \bibfield  {author} {\bibinfo {author} {\bibfnamefont {Y.}~\bibnamefont
  {Xie}}, \bibinfo {author} {\bibfnamefont {Y.}~\bibnamefont {Zhang}}, \bibinfo
  {author} {\bibfnamefont {M.}~\bibnamefont {Gong}}, \bibinfo {author}
  {\bibfnamefont {Z.}~\bibnamefont {Tang}},\ and\ \bibinfo {author}
  {\bibfnamefont {C.}~\bibnamefont {Han}},\ }\href@noop {} {\bibfield
  {journal} {\bibinfo  {journal} {Neural Networks}\ }\textbf {\bibinfo {volume}
  {132}},\ \bibinfo {pages} {180} (\bibinfo {year} {2020})}\BibitemShut
  {NoStop}%
\bibitem [{\citenamefont {Schmidt}\ \emph {et~al.}(2021)\citenamefont
  {Schmidt}, \citenamefont {Pettersson}, \citenamefont {Verdozzi},
  \citenamefont {Botti},\ and\ \citenamefont {Marques}}]{schmidt2021crystal}%
  \BibitemOpen
  \bibfield  {author} {\bibinfo {author} {\bibfnamefont {J.}~\bibnamefont
  {Schmidt}}, \bibinfo {author} {\bibfnamefont {L.}~\bibnamefont {Pettersson}},
  \bibinfo {author} {\bibfnamefont {C.}~\bibnamefont {Verdozzi}}, \bibinfo
  {author} {\bibfnamefont {S.}~\bibnamefont {Botti}},\ and\ \bibinfo {author}
  {\bibfnamefont {M.~A.}\ \bibnamefont {Marques}},\ }\href@noop {} {\bibfield
  {journal} {\bibinfo  {journal} {Science advances}\ }\textbf {\bibinfo
  {volume} {7}},\ \bibinfo {pages} {eabi7948} (\bibinfo {year}
  {2021})}\BibitemShut {NoStop}%
\bibitem [{\citenamefont {Banik}\ \emph {et~al.}(2023)\citenamefont {Banik},
  \citenamefont {Dhabal}, \citenamefont {Chan}, \citenamefont {Manna},
  \citenamefont {Cherukara}, \citenamefont {Molinero},\ and\ \citenamefont
  {Sankaranarayanan}}]{banik2023cegann}%
  \BibitemOpen
  \bibfield  {author} {\bibinfo {author} {\bibfnamefont {S.}~\bibnamefont
  {Banik}}, \bibinfo {author} {\bibfnamefont {D.}~\bibnamefont {Dhabal}},
  \bibinfo {author} {\bibfnamefont {H.}~\bibnamefont {Chan}}, \bibinfo {author}
  {\bibfnamefont {S.}~\bibnamefont {Manna}}, \bibinfo {author} {\bibfnamefont
  {M.}~\bibnamefont {Cherukara}}, \bibinfo {author} {\bibfnamefont
  {V.}~\bibnamefont {Molinero}},\ and\ \bibinfo {author} {\bibfnamefont
  {S.~K.}\ \bibnamefont {Sankaranarayanan}},\ }\href@noop {} {\bibfield
  {journal} {\bibinfo  {journal} {Npj Computational Materials}\ }\textbf
  {\bibinfo {volume} {9}},\ \bibinfo {pages} {23} (\bibinfo {year}
  {2023})}\BibitemShut {NoStop}%
\bibitem [{\citenamefont {Mo}\ \emph {et~al.}(2022)\citenamefont {Mo},
  \citenamefont {Huang}, \citenamefont {Xing},\ and\ \citenamefont
  {Lv}}]{mo2022multi}%
  \BibitemOpen
  \bibfield  {author} {\bibinfo {author} {\bibfnamefont {X.}~\bibnamefont
  {Mo}}, \bibinfo {author} {\bibfnamefont {Z.}~\bibnamefont {Huang}}, \bibinfo
  {author} {\bibfnamefont {Y.}~\bibnamefont {Xing}},\ and\ \bibinfo {author}
  {\bibfnamefont {C.}~\bibnamefont {Lv}},\ }\href@noop {} {\bibfield  {journal}
  {\bibinfo  {journal} {IEEE Transactions on Intelligent Transportation
  Systems}\ }\textbf {\bibinfo {volume} {23}},\ \bibinfo {pages} {9554}
  (\bibinfo {year} {2022})}\BibitemShut {NoStop}%
\bibitem [{\citenamefont {Zhang}\ and\ \citenamefont
  {Li}(2022)}]{zhang2022eesanet}%
  \BibitemOpen
  \bibfield  {author} {\bibinfo {author} {\bibfnamefont {J.}~\bibnamefont
  {Zhang}}\ and\ \bibinfo {author} {\bibfnamefont {Q.}~\bibnamefont {Li}},\
  }\href@noop {} {\bibfield  {journal} {\bibinfo  {journal} {Optics Express}\
  }\textbf {\bibinfo {volume} {30}},\ \bibinfo {pages} {10470} (\bibinfo {year}
  {2022})}\BibitemShut {NoStop}%
\end{thebibliography}%

\end{document}